\documentclass{article}
\usepackage{spconf,amsmath,graphicx}
\usepackage{hyperref}
\usepackage{amsmath}
\usepackage{mathptmx}
\usepackage{amsfonts}

\usepackage{enumitem}
\setlist{nosep, leftmargin=14pt}

\usepackage{mwe} 

\title{GM-LDM: Latent Diffusion Model for Brain Biomarker Identification through Functional Data-Driven Gray Matter Synthesis}
\name{Xu Hu\textsuperscript{*}\textsuperscript{1}\thanks{*These authors contributed equally}, Jingling Yang\textsuperscript{*}\textsuperscript{1}, Sihan Jia\textsuperscript{2}, Yuda Bi\textsuperscript{2}, and Vince D Calhoun\textsuperscript{2}}
\address{\textit{Anhui University, Hefei, China}\textsuperscript{1}\\\textit{TReNDS Center (GSU, Gatech, Emory), Atlanta, GA, USA}\textsuperscript{2}}
\begin{document}
%
\maketitle
\begin{abstract}

Generative models based on deep learning have shown significant potential in medical imaging, particularly for modality transformation and multimodal fusion in MRI-based brain imaging. This study introduces GM-LDM, a novel framework that leverages the latent diffusion model (LDM) to enhance the efficiency and precision of MRI generation tasks. GM-LDM integrates a 3D autoencoder, pre-trained on the large-scale ABCD MRI dataset, achieving statistical consistency through KL divergence loss. We employ a Vision Transformer (ViT)-based encoder-decoder as the denoising network to optimize generation quality. The framework flexibly incorporates conditional data, such as functional network connectivity (FNC) data, enabling personalized brain imaging, biomarker identification, and functional-to-structural information translation for brain diseases like schizophrenia.

\end{abstract}
\begin{keywords}
Magnetic resonance imaging, latent diffusion model, vision transformer, generative model
\end{keywords}
\section{Introduction}
\label{sec:intro}

Generative models have transformed medical imaging by enabling advanced applications in MRI-based neuroimaging \cite{BI2024120674}. Models such as generative adversarial networks (GANs) \cite{goodfellow2014generative} and diffusion models \cite{ho2020denoising} excel in tasks like modality fusion (e.g., combining T1- and T2-weighted images) and clinical applications, such as MRI-to-CT conversion, which reduces the need for radiation-based CT scans \cite{zhuang2024segmentation}. Despite their strengths, GANs face challenges like mode collapse and unstable training \cite{ferreira2024gan}, while diffusion models, particularly latent diffusion models (LDMs) \cite{rombach2022high}, address these issues by mapping images to a latent space via pre-trained autoencoders. However, in MRI imaging, data scarcity limits the pre-training of autoencoders, hindering the full potential of LDMs \cite{zhang2024phy, pinaya2022brain, fan2024survey, cao2024survey}.

\begin{figure}
    \centering
    \includegraphics[width=1\linewidth]{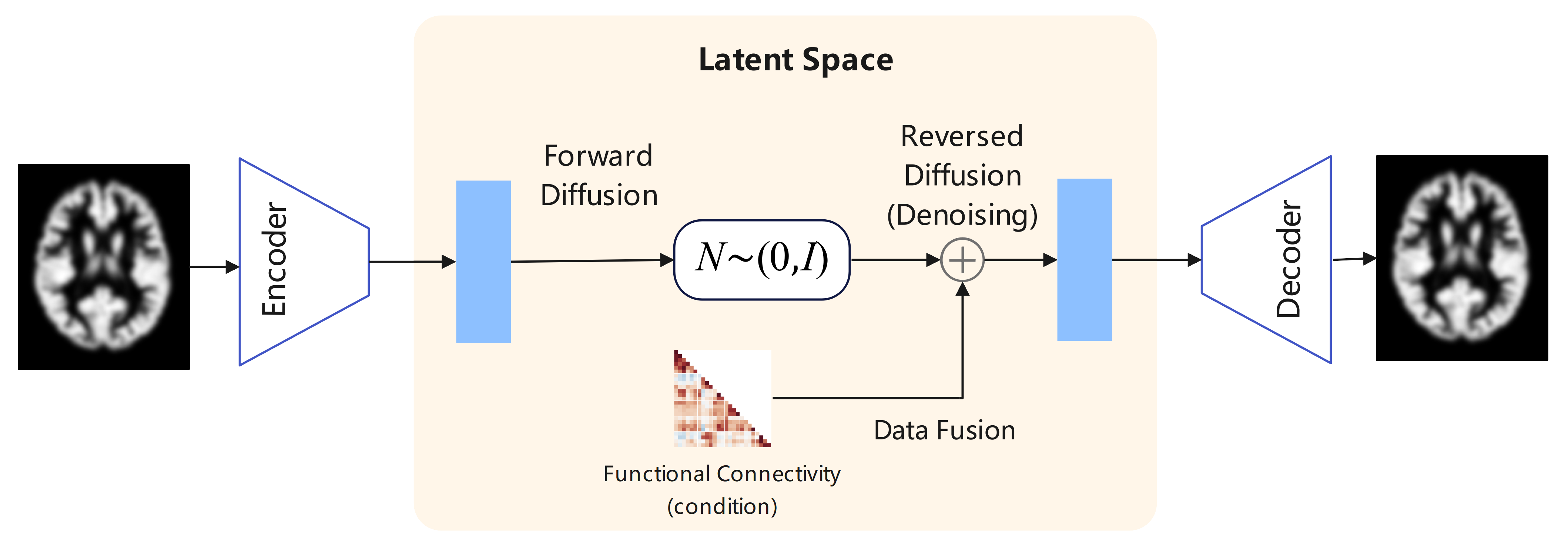}
    \caption{The overall pipeline of our latent diffusion model guided with conditions}
    \label{fig:enter-label}
\end{figure}

\begin{figure*}
    \centering
    \includegraphics[width=0.65\linewidth]{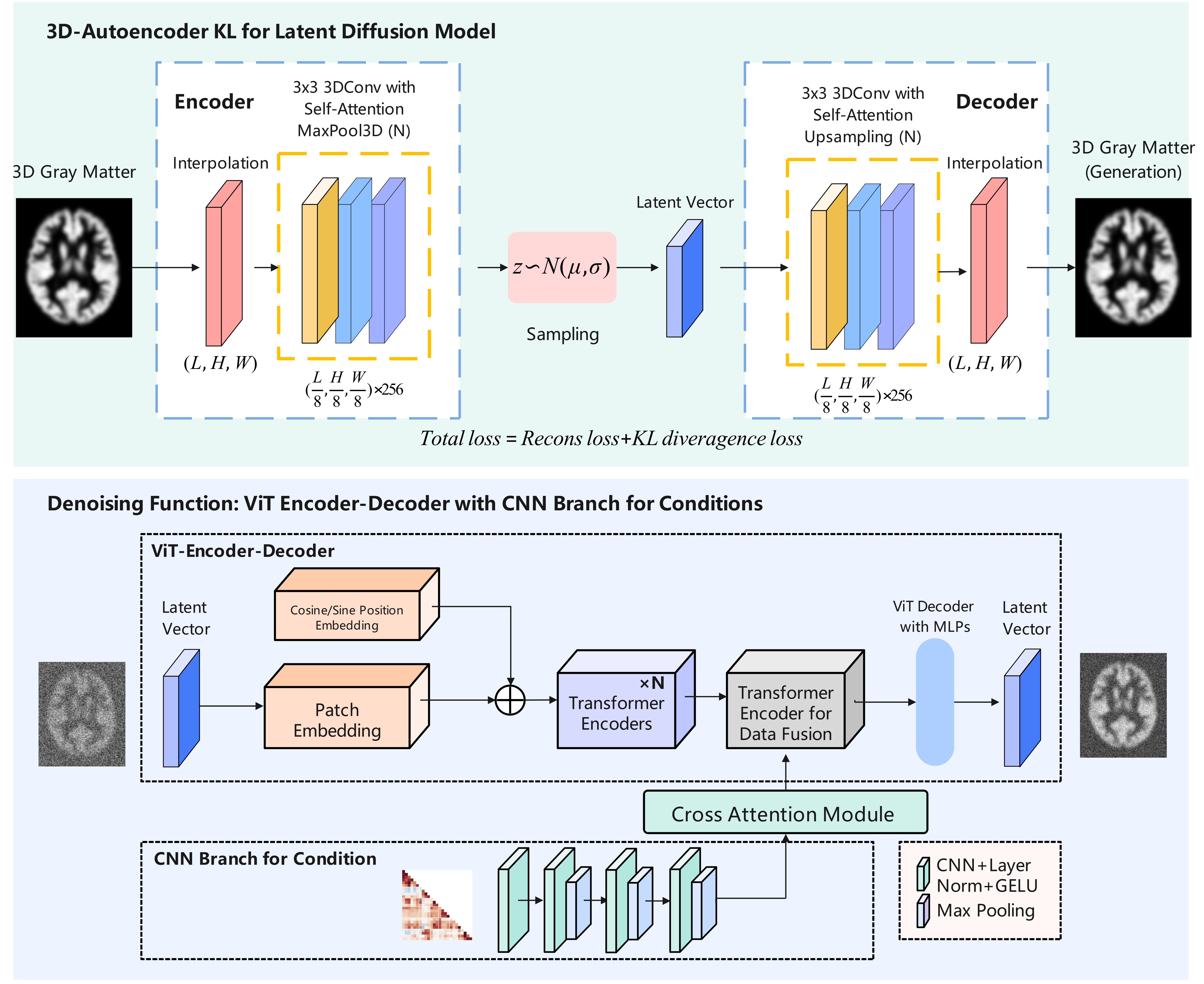}
    \caption{The architectures of 3D-autoencoder with kl divergence loss and multimodal denoising network via ViT, CNN, and cross-attention module. }
    \label{fig:enter-label}
\end{figure*}

To overcome these challenges, we propose GM-LDM, a framework designed to advance latent diffusion model (LDM) applications in medical imaging. Our contributions are threefold: (1) We developed a 3D gray matter (GM) generation framework using an LDM with a 3D autoencoder pre-trained on massive 3D MRI datasets, including the Adolescent Brain Cognitive Development (ABCD) study and UK Biobank. This autoencoder serves as a powerful representation learning tool, leveraging large-scale training to efficiently extract GM features and map 3D MRI data to a reduced latent space with high fidelity. It acts as a benchmark for downstream tasks, with learnable interpolation layers enhancing its adaptability to diverse medical image shapes and modalities. (2) We introduced a hybrid denoising framework combining convolutional neural networks (CNNs) and a Vision Transformer (ViT)-based encoder-decoder \cite{dosovitskiy2020image}. The ViT backbone integrates CNN-extracted state features at various stages, significantly improving resolution and modality transformation accuracy. (3) We validated the framework by generating subject-specific 3D GM images conditioned on functional network connectivity (FNC) data, confirming schizophrenia-related biomarkers. These results highlight GM-LDM’s potential for disease diagnosis and target localization, offering a robust tool for clinical neuroscience.

\section{Related Works}

Jiang et al. \cite{jiang2023cola} introduce a conditioned LDM for multi-modal MRI synthesis, operating in latent space to reduce memory usage, with structural guidance via brain region masks to maintain anatomical details. Kim et al. \cite{kim2024adaptive} propose an adaptive LDM (ALDM) for 3D MRI translation, enabling multi-modal translations from a single source, outperforming other models. Pan et al. \cite{pan2024synthetic} present MRI-to-CT denoising diffusion (MC-DDPM), using a Swin-Vnet-based reverse diffusion process to generate high-quality synthetic CT images from MRI data. In reviewing recent studies, we find that while these models introduce innovative architectures, certain limitations remain. For example, the lack of a portable autoencoder for large-scale training and the limited exploration of links between generative models and biomarker discovery represent areas for further improvement.

\section{Methods}

\subsection{Latent Diffusion Models}
Latent diffusion models (LDMs) generate high-quality images efficiently, making them well-suited for MRI-based neuroimaging tasks like modality transformation and multimodal fusion. Let \(\mathbf{x}_0 \in \mathcal{X}\) represent a 3D MRI volume in the data space \(\mathcal{X}\), with latent representation \(\mathbf{z}_0 \in \mathcal{Z}\) obtained via an encoding function \(E: \mathcal{X} \to \mathcal{Z}\). The latent diffusion process adds noise to \(\mathbf{z}_0\) over steps \(\mathbf{z}_t\), \(t = 0, 1, \dots, T\), through a Markov process. The forward diffusion is:

\[
q(\mathbf{z}_t | \mathbf{z}_{t-1}) = \mathcal{N}(\mathbf{z}_t; \sqrt{1 - \beta_t} \mathbf{z}_{t-1}, \beta_t \mathbf{I}),
\]

where \(\beta_t\) is a variance schedule controlling noise. The reverse process recovers \(\mathbf{z}_0\) from \(\mathbf{z}_T\) using a parameterized denoising model \(p_\theta(\mathbf{z}_{t-1} | \mathbf{z}_t, \mathbf{c})\), where \(\mathbf{c}\) is conditional information, such as FNC data, guiding modality-specific generation:

\[
p_\theta(\mathbf{z}_{t-1} | \mathbf{z}_t, \mathbf{c}) = \mathcal{N}(\mathbf{z}_{t-1}; \mu_\theta(\mathbf{z}_t, \mathbf{c}, t), \Sigma_\theta(\mathbf{z}_t, t)),
\]

with \(\mu_\theta\) and \(\Sigma_\theta\) as learned mean and variance functions. The denoised \(\mathbf{z}_0\) is decoded via \(D: \mathcal{Z} \to \mathcal{X}\), yielding \(\hat{\mathbf{x}}_0 = D(\mathbf{z}_0)\). The process, shown in \textbf{Fig. 1}, is:

\[
\mathbf{x}_0 \xrightarrow{E} \mathbf{z}_0 \xrightarrow{\text{diffusion}} \mathbf{z}_T \xrightarrow{\text{denoising} \ \mathbf{c}} \mathbf{z}_0 \xrightarrow{D} \hat{\mathbf{x}}_0(\mathbf{c}).
\]

This framework enables efficient generation of subject-specific MRI images conditioned on FNC data, supporting biomarker identification for brain diseases.

\subsection{3D Autoencoder}
The 3D autoencoder is a core component of the GM-LDM framework, designed to learn robust representations of 3D MRI data for efficient mapping to the latent space. Pre-trained on large-scale 3D MRI datasets, the autoencoder captures GM features critical for downstream neuroimaging tasks. Let the input be a 3D volume \(\mathbf{x} \in \mathbb{R}^{L' \times W' \times H'}\), where \(L', W', H'\) denote the initial spatial dimensions. The input is first processed by a learnable interpolation layer \(\mathcal{I}_1: \mathbb{R}^{L' \times W' \times H'} \to \mathbb{R}^{L \times W \times H}\), which adapts the input to a standardized shape \((L, W, H)\) to accommodate varying MRI resolutions and modalities.

The encoder network \(E\) transforms the interpolated input into a latent representation \(\mathbf{z}_0 \in \mathbb{R}^{256 \times \frac{L}{8} \times \frac{W}{8} \times \frac{H}{8}}\). This representation is modeled as a Gaussian distribution \(\mathcal{N}(\mu_E(\mathbf{x}), \sigma_E(\mathbf{x}))\), where \(\mu_E(\mathbf{x})\) and \(\sigma_E(\mathbf{x})\) are the mean and variance learned by the encoder. The latent representation is sampled as:

\[
\mathbf{z}_0 = \mu_E(\mathbf{x}) + \sigma_E(\mathbf{x}) \cdot \epsilon, \quad \epsilon \sim \mathcal{N}(0, I).
\]

The decoder network \(D\) reconstructs the latent representation \(\mathbf{z}_0\) into a feature map of shape \(\mathbb{R}^{256 \times \frac{L}{8} \times \frac{W}{8} \times \frac{H}{8}}\). This is upsampled to \((L, W, H)\) using an upsampling function \(\mathcal{U}: \mathbb{R}^{256 \times \frac{L}{8} \times \frac{W}{8} \times \frac{H}{8}} \to \mathbb{R}^{L \times W \times H}\). A second learnable interpolation layer \(\mathcal{I}_2: \mathbb{R}^{L \times W \times H} \to \mathbb{R}^{L' \times W' \times H'}\) restores the output to the original input shape, yielding:

\[
\hat{\mathbf{x}} = \mathcal{I}_2(\mathcal{U}(D(\mathbf{z}_0))) \in \mathbb{R}^{L' \times W' \times H'}.
\]

The interpolation layers \(\mathcal{I}_1\) and \(\mathcal{I}_2\) ensure compatibility across diverse MRI datasets, enhancing the autoencoder’s flexibility for tasks like cross-modality generation.

\subsection{Loss Functions}
The 3D autoencoder is trained using a combination of KL divergence and reconstruction losses to ensure statistical consistency in the latent space and accurate image reconstruction. The KL divergence loss (\(\mathcal{L}_{\text{KL}}\)) measures the difference between the learned latent distribution \(q(\mathbf{z}_0|\mathbf{x})\) and the prior distribution \(p(\mathbf{z}_0) = \mathcal{N}(0, I)\), defined as:

\[
\mathcal{L}_{\text{KL}} = D_{\text{KL}}(q(\mathbf{z}_0|\mathbf{x}) \parallel p(\mathbf{z}_0)) = \mathbb{E}_{q(\mathbf{z}_0|\mathbf{x})} \left[ \log \frac{q(\mathbf{z}_0|\mathbf{x})}{p(\mathbf{z}_0)} \right].
\]

For a Gaussian posterior \(q(\mathbf{z}_0|\mathbf{x}) = \mathcal{N}(\mu_E(\mathbf{x}), \sigma_E(\mathbf{x}))\), the KL divergence is computed analytically as:

\[
\mathcal{L}_{\text{KL}} = -\frac{1}{2} \sum_{i=1}^{d} \left( 1 + \log \sigma_E^2(\mathbf{x}_i) - \mu_E^2(\mathbf{x}_i) - \sigma_E^2(\mathbf{x}_i) \right),
\]

where \(d\) is the dimensionality of the latent space. This loss regularizes the latent space to align with the prior distribution. The reconstruction loss (\(\mathcal{L}_{\text{recon}}\)) quantifies the fidelity of the reconstructed output \(\hat{\mathbf{x}}\) compared to the input \(\mathbf{x}\):

\[
\mathcal{L}_{\text{recon}} = \frac{1}{N} \sum_{i=1}^{N} (\mathbf{x}_i - \hat{\mathbf{x}}_i)^2,
\]

where \(N\) is the number of data points. The total loss combines these terms, weighted by a hyperparameter \(\alpha \in [0, 1]\):

\[
\mathcal{L}_{\text{total}} = \alpha \mathcal{L}_{\text{KL}} + (1 - \alpha) \mathcal{L}_{\text{recon}}.
\]

The hyperparameter \(\alpha\) balances the trade-off between latent space regularization and reconstruction fidelity, tuned empirically to optimize performance for MRI data.

\subsection{Denoising Network}
The denoising network is a critical component of the GM-LDM framework, designed to recover \(\mathbf{z}_0\) from the noisy latent variable \(\mathbf{z}_T\) by integrating conditional information. It employs a hybrid architecture combining a Vision Transformer (ViT)-based encoder-decoder with CNNs. The encoder consists of 12 transformer layers, each applying self-attention and a feed-forward network to process the noisy latent representation \(\mathbf{z}_T\).

The decoder reconstructs \(\mathbf{z}_0\) by incorporating conditional information \(\mathbf{c}\), representing FNC data, extracted using a CNN. The CNN extracts multi-scale features from \(\mathbf{c}\), which are fused into the ViT decoder via a cross-attention (CA) mechanism. The cross-attention layer is formulated as:

\[
\text{CA}(\mathbf{Q}, \mathbf{K}_{\text{cond}}, \mathbf{V}_{\text{cond}}) = \text{softmax}\left( \frac{\mathbf{Q} \mathbf{K}_{\text{cond}}^T}{\sqrt{d_k}} \right) \mathbf{V}_{\text{cond}},
\]

where \(\mathbf{Q}\) is derived from the decoder’s input, and \(\mathbf{K}_{\text{cond}}\) and \(\mathbf{V}_{\text{cond}}\) are computed from the CNN-extracted features of \(\mathbf{c}\). This is followed by parallel multi-layer perceptrons (MLPs) to finalize the reconstruction, inspired by \cite{lee2021vitgan}. 

\section{Experiments and Results}

\begin{table*}[ht]
\centering
\caption{Model performance details for baselines, comparisons, and our preferred model.}
\begin{tabular}{llllll}
\hline
\textbf{Name} & \textbf{Model} & \textbf{Pre-trained} & \textbf{Condition} & \textbf{Pearson Corr} & \textbf{SSIM} \\ \hline
Baseline-1   & Diffusion      & No        & Random-vector & 0.78 & 0.77 \\
Baseline-2   & LDM            & No        & Random-vector & 0.79 & 0.79 \\
Comparison-1 & LDM            & Yes       & Random-vector & 0.86 & 0.84 \\
Comparison-2 & LDM            & No        & FNC           & 0.83 & 0.82 \\
GM-LDM       & LDM            & Yes       & FNC           & 0.89 & 0.86 \\ \hline
\end{tabular}
\end{table*}

\subsection{Datasets}

We used the large-scale \href{https://abcdstudy.org/}{ABCD dataset} (n=11,220) to train our 3D autoencoder. The ABCD dataset contains comprehensive neuroimaging and cognitive data from adolescents, providing a rich source for evaluating brain structure and function through MRI scans. Additionally, we leveraged the \href{https://www.ukbiobank.ac.uk/}{UK Biobank} dataset, which includes over 40000 MRI scans, offering a diverse and extensive resource for pre-training our autoencoder to capture robust brain imaging features. For brain disease analysis, we employed two schizophrenia-related datasets, combining data from three international studies (fBIRN, MPRC, COBRE) and several hospitals in China. The dataset included 1,642 participants, with 803 healthy controls and 839 individuals with schizophrenia. Resting-state fMRI data were collected using 3.0 Tesla scanners across multiple sites, with EPI sequences (TR/TE ~2000/30 ms, voxel sizes from 3 × 3 × 3 mm to 3.75 × 3.75 × 4.5 mm).

\subsection{Model Design, Training, and Validation}

The 3D autoencoder, a cornerstone of the GM-LDM framework, employs an encoder-decoder architecture with eight convolutional layers per module, augmented by self-attention mechanisms to capture long-range dependencies in 3D MRI data. It was pre-trained and fine-tuned on the ABCD and UkBiobank dataset to learn robust GM representations, using fp16 precision for computational efficiency, a MultiStepLR learning rate scheduler, and the AdamW optimizer (learning rate: 2e-4, batch size: 16). Training was conducted on four NVIDIA H100 GPUs to accelerate convergence. For the GM-LDM model, we focused on training the denoising network, which integrates a Vision Transformer (ViT)-based encoder-decoder with convolutional feature extraction, to generate subject-specific GM images conditioned on FNC data. The GM-LDM adopted a similar training configuration, but with a reduced learning rate of 3e-5 to ensure stable denoising, and utilized four H100 GPUs. To enhance robustness and generalization, we applied 5-fold cross-validation during training. To evaluate model performance, we conducted baseline and comparative experiments, including training the 3D autoencoder without ABCD pre-training and using a standard diffusion model without latent space mapping. The similarity between generated and real GM images was assessed using Pearson correlation, chosen for its sensitivity to the structural and grayscale properties of MRI data. Additionally, we compared the effectiveness of FNC data as a conditioning factor against random noise of equivalent dimensions, confirming FNC’s role in guiding accurate image generation.

\subsection{Basic Results}

\textbf{Table 1} summarizes the performance of our GM-LDM model against baselines and comparisons. GM-LDM, conditioned on schizophrenia-related FNC data with pre-training, achieved the highest Pearson correlation (0.89) and SSIM (0.86), outperforming models using random-vector conditioning or lacking pre-training. Baseline-1 (Diffusion, no pre-training, random-vector) and Baseline-2 (LDM, no pre-training, random-vector) scored 0.78/0.77 and 0.79/0.79 (Pearson/SSIM), respectively, showing limited effectiveness. Comparison-1 (LDM, pre-trained, random-vector) reached 0.86/0.84, highlighting pre-training benefits, while Comparison-2 (LDM, no pre-training, FNC) scored 0.83/0.82, emphasizing FNC’s role. These results confirm that pre-training on large datasets (e.g., ABCD) and FNC conditioning enhance schizophrenia-related GM generation.

\begin{figure}
    \centering
    \includegraphics[width=1\linewidth]{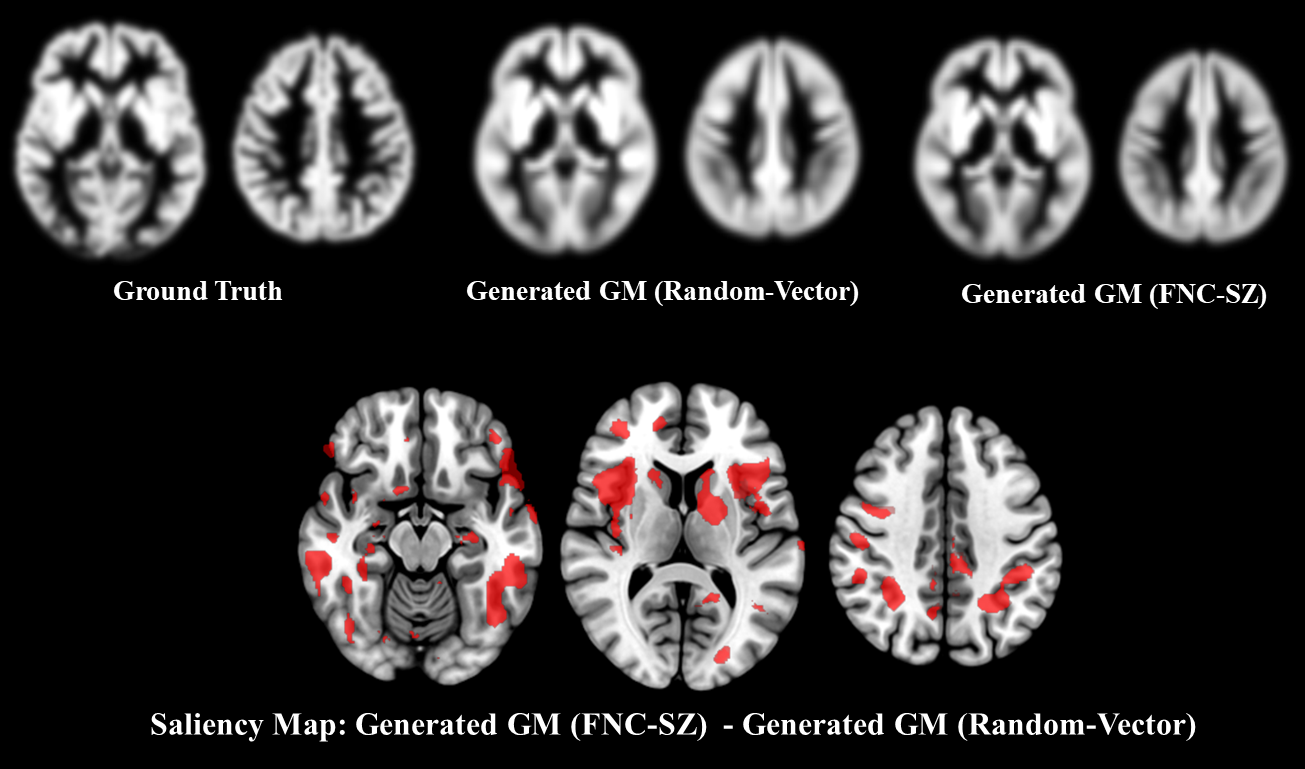}
    \caption{This figure shows GM examples generated from both random vectors and schizophrenia-related FNC data, emphasizing differences between the two GMs.}
    \label{fig:enter-label}
\end{figure}

\subsection{Biomarkers Discovery}

By comparing GM generated under the guidance of FNC data from subjects with schizophrenia to GM generated with random vector guidance, we observed that certain brain regions were more prominent in the FNC-guided GM. \textbf{Fig. 3} illustrates the generated GM and highlights brain regions associated with schizophrenia. Analysis of these regions revealed similarities with biomarkers identified in our previous research \cite{BI2024120674}, where GM served as a conditional input to generate FNC matrices for schizophrenia subjects, and attention weights were applied to create GM saliency maps. These saliency maps emphasized critical brain regions, providing insights into the spatial importance of GM structures contributing to FNC patterns in schizophrenia. In this study, we identified a strong link between the cerebellum and schizophrenia, consistent with traditional research demonstrating cerebellar involvement in cognitive dysmetria in schizophrenia \cite{andreasen1996schizophrenia}. Additionally, we observed significant associations between schizophrenia and specific basal ganglia structures, such as the caudate and putamen, which have been implicated in cognition and emotion in prior studies \cite{mamah2007structural}. These findings align with established literature and underscore the robustness of our approach. Our model demonstrates potential for broader application in identifying regions of interest (ROIs) associated with various brain disorders.

\section{Conclusions}

Our model, based on a basic model pre-trained on a large dataset, has achieved strong performance in research focused on brain disorders using smaller, disease-specific datasets. In the future, our model will be widely applied to explore and validate biomarkers associated with various brain disorders. These novel or potential biomarkers hold great promise for advancing our understanding of brain disorders and the structural characteristics unique to specific individuals, providing valuable insights into the pathology of neurological diseases.

\bibliographystyle{IEEEbib}
\bibliography{strings,refs}

\end{document}